\documentclass[a4paper,12pt] {article}
\usepackage{latexsym}
\makeatletter

\makeatletter
 \@addtoreset{equation}{section}

 \makeatother

\begin{document}

 \normalsize
 \title{The one example of  Lorentz group}

\bigskip
\author{Leonid D. Lantsman, \\
{\it Wissenschaftliche Gesellschaft} {\rm ZWST},
 {\it Judische Gemeinde  zu Rostock,} \\
 Wilhelm-K$\ddot u$lz Platz,6. 18055 Rostock.\\
e-mail:  llantsman@freenet.de}

\date{}

\maketitle

\abstract {
The aim of this work is to show, on  the example of the behaviour
of the spinless charged particle  in the homogeneous electric
field ,that one can  quantized   the velocity of particle  by the
special gauge fixation. The work gives  also the some information
about the theory  of second quantisation in the space of
Hilbert-Fock and the theory of projectors in the Hilbert space.
One consider in Appendix the  theory of the spinless charged
particle in the homogeneous addiabatical changed electrical
field.}

\thispagestyle{empty}
\renewcommand{\thefootnote}{\arabic{footnote}}
\newpage
\setcounter{page}1

\section* {1.Introduction}
The history of   physics in xx century is determined with the two great discoveries. This is the  A. Einstein's relativity theory
(the special and the general relativity) and the quantum mechanics.The hypothesis about the  discrete nature of light,which
was suggested by M.K.E.L.Planck in 1900y.(as a 'saving' from the \it  ultra-violet catastrophe\rm:the
 endless character of  the energy density of the eradiation spectrum
because of the contribution of the ultra-violet part of
the spectrum ) led M.Planck ( in fact against his 'classical education' as a physicist of xix century with its Maxwell electrodynamics and Newton-Lagrange mechanics) to the conception of  the \it quantum of minimal\\ action
\rm,i.e. to the Planck constant $\hbar$  .This was a greatest drama of doctor's Planck life: he could not accept up to his death in 1947 y. the 'quantum revolution'.\\What is this- the quantum of minimal action? Our reader ,which acquainted with classical Lagrange  mechanics, is knows
that [1] one can represent  the phase space  $ \Gamma$
  as a \\\
\begin{displaymath}
 \Gamma =\int dq_ 1...dq_ sdp_1...dp_s\ (1)
\end{displaymath}
 where $q_i,p_i$  are the canonical co-ordinates and the canonical momentum correspondingly.
The M.Planck's hypothesis, in the terms of formula (1),signifies that \it the cell of the phase space with the size $\Gamma\leq{(2\pi\hbar)}^3$\it  is not exists \rm! This conclusion(at least for the quasiclassical  approach in quantum mechanics)
 signifies non other then  \it Heisenberg inequality\rm -the fundamental low of quantum mechanics!So,on  author's of this article
opinion,an opportunity for  the quantum mechanics appearance was as long ago in 1900 y.(with account of the accumulated knowledge in geometrical optics
:the Fermat principle,-the low of minimal action in geometrical optics, and the optical-mechanical analogy[2].In fact,one can interpret  geometrical optic \it as a quasiclassic theory of photon \rm!)
\\ Albert Einstein as it well known also could not accept  quantum mechanics (by  the of many years friendship with N.Bohr.
 Their discussions, the examples and  the counterexamples 'for' and 'against' of
the quantum theory,-all this had often the highly  stormy nature!)
\\Indeed,the 'precipice' between these two great theories is not large.There are
 (quantum mechanics and the  relativity theory)- the two corner-stones,on whichis based the majestic building of  modern theoretical physics.
 One of examples -the quantum -field theories:quantum electrodynamics,QCD, the electroweak theory.They all are the relativistic quantum theories.
\\The basic postulates of the quantum -field theory,-  \it  Wigthman axioms \rm [3] , are based on
Lorentz-Poincare group in Minkowski space
.The demand of Poincare invariantness of \it Wigthman functions\rm (the vacuum averages of the operator product of some quantum-fields),
i.e.the  demand of relativistic invariantness of S- matrix and  the effective section; \it the microcause  principle\rm,i.e.  zero value of the fields (anti)commutator at the space-like interval between them ,-
there are the basic displays  of special relativity in the quantum -field theory.
\\The cause of such large difference between quantum mechanics and  'old'  Lagrange-Newton  (classical)
  mechanics is the Planck constant $\hbar$. \rm The introduction of the quantum  $\hbar$
in  theoretical physics  lead to the replacement of  \it Poisson brackets\rm onto  the \it commutation \rm relation between
 the canonical co-ordinates and the canonical momentum.
These \it canonical comutation relations \rm turn  into  Poisson brackets of Hamilton mechanics in the
 limit of  $\hbar$=0
[4]. \bf The construction of the eigenfunctions of the quantum operators (i.e. of the vectors of Hilbert
space):in particular for the quantum-field operators, which we consider as a canonical  co-ordinates or a
 canonical momentum; the solution of the movement equations \\(Schrodinger equation in non-relativistic
 quantum mechanics or the quantum-field equations (Dirac equation,Klein-Gordan-Fock equation,
Rarita-Schwinger equation,- are the basic examples of the relativistic equations),- are the basic problems
of the quantum theory)\rm  . This is very important that the quantum- field operators have the dimension
of the canonical co-ordinates or the canonical momentum.\\ One writes  down the movement equations,
issuing from the classical Lagrangian of corresponding particle and the Lagrange equations as a
conditions of the minimal action  (for example- the case of the scalar particle).Then we replace  all  \it
classical \rm   momenta on the momentum operators $-i\hbar\partial/\partial x_i$. One calls all this
 process as the \it  first quantization \rm   (it is appropriate to mention here that the conclusion of Dirac
equation is based on  Schrodinger equation and on the superposition principle ( the basic
principle of  quantum mechanics) which dictates the form of energy operator (the quantum-mechanical
generalization of Hamiltonian):this operator is proportional to  Dirac matrixes[5].Thus we obtain this
equation from the \it quantum-mechanical principles\rm . But we can conclude it as  Lagrange equation
[6]  ).
\\\it The second quantization \rm this is ([7],[3]) the  point of view onto the quantum field as a
 \it multi-particle field \rm .The every particle we then interpret as a \it one degree of freedom \rm .
Then we construct   \it Hilbert-Fock space of second quantization\rm  .
\\The form of the  eigenvectors of Hilbert-Fock space is given, for example, in monograph [3]
 (look formula (7.99) ).This form depends on helisity s (explicity and through the spinors
 $\omega_{\alpha i} $. On the language of Penrose \it flag \rm  structure of the space-time, the  spinors
$\omega_{\alpha i} $ are [8]  \it the main spinors \rm ,by which we decompose  arbitrary spin-tensor
  \\$ \psi( \omega,\varpi ) =\sum_{\alpha 1...\alpha 2j;\beta 1'...\beta 2k'} \omega^{\alpha1}...\omega^{\alpha2j} \varpi ^{\beta 1'}...\varpi^{\beta 2k'} $
\ (2)
\\Thus  formula (7.99) in[3] is in fact the usual decomposition of the wave function on the space and  the
 spinor parts ( in the  momentum representation)
\\The structure of Hilbert space is determined with[3]  \it  Gelfand-Naimark-Segal \\\rm (\it GNS\rm)\it
construction.
 \rm Let some algebra with involution U is given. Let us denote this algebra as C*.Then  exists the
 isomorphism $ \pi$ of algebra U into  algebra B (\it H\rm) of all linear limited operators in the Hilbert
 space.The representation $ \pi$ is called \it unreducible \rm if  every closed subspace in \it H\rm , which
is invariant relatively to all operators  $\pi (A); A\in U $ is
 $\emptyset$ or all \it H\rm.The vector $\Phi \in \it H $ \rm  is called  \it the cyclical vector \rm for  representation $ \pi$ if all vectors of the form $\pi (A) \Phi$ where $ A \in U
 $ form the total set in \it H\rm
 ( correspondingly , the such representation with the cyclical vector is called \it cyclical\rm  ).
\\If $\Phi$ is  the vector in  \it H \rm  then it generates the \it positive functional
\\$ F_\Phi =< \Phi ,\pi (A)  \Phi >$ \rm \ (3) \\ on U (in the terms of the  probability theory it is the
 \it mathematical expectation \rm  of the value $\pi (A)$   \it in  state  $\Phi$ \rm ).This functional is called
  \it the vector\\ functional , associated with  representation  $\pi $ \it  and   vector  $\Phi$.
\\ \rm  In these terms  GNS construction consists in following:\\\it One can determine some (cyclical)
 representation  $\pi_F$  \it of algebra \rm U \it in the Hilbert space with cyclical   vector  $\Phi_F$ \it  for
 given positive functional \rm F  \it such that
  \\$ F(A)= <\Phi_F, \pi_F (A) \Phi_F >$ \rm \  (4)\\ \it The representation   $\pi_F$ \it is determined with
 these conditions unique with precision of the unitary equivalence.
\\ \rm The one of the most important for the quantum-field theory are  \it  birth and annihilation
operators.
\\ \rm  As it is well known, the spin of the particle determines the  two kinds of statistics : \it  the
 Bose-Einstein statistics for the integer spin and the Fermi statistics for the semi-integer spin \rm (the
  Maxwell statistics is the limit of these  statistics by $\hbar \to 0 $ ). This two kinds of statistics
 determine the  two kinds of the commutation relations  between the birth-annihilation operators
( we must consider  the commutation relations  for the  Bose  statistics and the anticommutation relations
 for the  Fermi  statistics). The every physicist-theorist ought to known these relations ,therefore one can
 omit they here. But following is most important: the action of the birth and  the annihilation  operators
onto  the cyclical vector ( in theoretical physics the cyclical vector is called  \it the vacuum vector\rm) .As it is usual in physics, let us denote the  vacuum vector as $\vert 0 >$  (\it the  'bra' vector\rm). Then
the  action of the  annihilation and the birth operators on $\vert 0 >$ is expressed as
\\ \begin{equation} a ( \Phi ) \vert 0 > =0\\  a*(
 \Phi ) \vert 0 > =\Phi  
\end{equation} \   (5) \\ for the birth operators  $ a^*(\Phi ) $ and the  annihilation operators
  $ a(\Phi ) $.\\ The relation (2) and formula (7.99) in [3] show us  that one can interpret the vectors of the
 Hilbert-Fock space as a \it  tensor products\rm : there are the  \it  tensor functions of such number of
 momentum ,how much the particles we consider in our model.
\rm But one can decompose the every tensor onto  \it the symmetrical and the antisymetrical parts.
 \rm And what is more:as we  know, the  statistics determine the form of this tensor [4]: some symmetrical
 wave function for Bose statistics and some antisymmetrical wave function for Fermi statistics (it is in
 fact [8] the consequence of  flag structure  and  spinor algebra,based on this structure).
\\ If we denote the one-particle  Hilbert space as a $\Lambda_1$, then one can determine by the natural
way the multiparticle space   as
 $\Lambda_1^{\otimes_ n} $ ,- \it the $n^{th}$\it  (anti)symmetrical  tensor power  of the  Hilbert space
$\Lambda_1$,\rm which we will denote as  $\Lambda_1^{\vee n} $ for Bose statistics and
 $\Lambda_1^{\wedge n} $ for Fermi statistics.
 As a particular case, by n=0 we have the scalar field with the usual scalar product $ \Phi_0 \Psi_0 $
\\ The vectors of the  states with different numbers of particles form the  mutually-orthogonal subspaces
 in the complete Hilbert  space. Hence, \it it should to introduce the direct sum of the $n$-\it particle
subspaces.\rm The direct sum  \\$\bigoplus_ {n=0}^\infty \Lambda_1^{\otimes_ n} $\ (6)\\ is called
\it the tensor algebra over the Hilbert space  $\Lambda_1$. \rm The vectors  of this space are the arbitrary
sequences $ \lbrace\Phi_n \rbrace_{n=0}^\infty $  such that  $\Phi_n \in   \Lambda_1^{\otimes_ n} $
and \\$ \vert\vert \Phi  \vert\vert^2 <\infty $ \   (7)\\
 and the scalar product has the form \\$ < \Phi, \Psi > = \sum _ {n=0}^\infty <\Phi_n , \Psi_n > $ \ (8)
 \\ It is obvious that one can identify  $ \Lambda_1^{\otimes_ n} $  with the n-particle space in above
direct sum. If  k $\neq$ n  then  $\Phi_k =0$. In particular,the  one-dimensional 0-particle  space
$  \Lambda_1^{\otimes_ 0} $ is called  \it the vacuum space. \rm It is \it  'pulled'  \rm onto  the vector
 $\Phi_0 \equiv \vert 0> $ with its components
\\$\ (\Phi_0)_0 =1;    (\Phi_0)_n =1$ \ by n$\neq0 $ \    (9) \\which is called  \it the vacuum vector. \rm
We can consider $\Phi_n$  as a projection of the sequence $\Phi$ on $ \Lambda_1^{\otimes_ n} $.
The vector $\Phi$  is called \it the  finite vector, \rm if it has\it  the finite number of its projections
 $\Phi_n$ \it different from zero.\rm  It is obvious that the finite vectors form the linear manifold which is
\it dense\rm  in the direct sum (6).\\ The direct sum (6) is divided with the natural way on the direct sum
\ it of its Bose and Fermi components, \rm which we will denote as
 $ F_\vee (\Lambda_1)$ and
$ F_\wedge (\Lambda_1) $ correspondingly.These spaces are called  \it the Fock spaces of bosons and
fermions \rm  correspondingly.\\ The fundamental quantum-mechanical characteristics of the particle are
\it  its mass and its spin.\rm We shall consider the one-particle space $\Lambda_1$ as a 'supplied'
with these  characteristics, and shall denote it as $\Lambda^ {(m,s)}$ .
This space is transformed by the unitary representation of Poincare group ( latter acts as an
\it automorphism \rm on $\Lambda^ {(m,s)}$  ).\\ One can,of course, consider the  spaces
$\Lambda_1^{\vee n} $  and  $\Lambda_1^{\wedge n} $  as an  \it eigen-subspaces of the
 (Hermit) operator of the numbers of particles\rm N, which is determined on the (finite) vectors by
 formula\\$ (N \Phi)_n = n \Phi_n $ \ (10) \\ \it The construction of Fock space  $ F_\vee (\Lambda_1)$  \it or
 $ F_\wedge (\Lambda_1)$ \it  by given space $\Lambda^{(m,s)} $  \it is called the second quantization\\ \rm The  birth and the annihilation  operators allow the interpretation  in the terms of the number of particles.
 Let us determine the symmetrical tensor product of vectors  $\Phi, \Psi$ of  space
$ F_\vee (\Lambda_1)$ which is  \it associative,distributive and commutative.\rm And analogous
 we determine the antisymmetrical  tensor product of vectors  $\Phi, \Psi$ of  space
 $ F_\wedge (\Lambda_1)$ which is  \it  distributive and associative only.
\\$ \Phi \vee \Psi = $ \rm Sym $ \Phi\bigotimes \Psi $ \\$\Phi \wedge \Psi =
$ \rm Antisym $ \Phi\bigotimes \Psi $ \rm \ (11)\\  Let us fix  some vector $\Phi$ in formula (11) and let us
 consider then the maps  \\$ \Psi \to \sqrt N \Phi \vee \Psi $\  (12a) \\or \\
$ \Psi \to \sqrt N \Phi \wedge \Psi $\ (12b)
\\ The both maps are some linear operators on the finite vectors  $\Phi$ of Fock (boson,fermion) spaces.
 The basic feature of these maps is that  \it the \rm  n- \it particle vector \it turns into the \rm  (n+1)-\it particle
 vector , \rm therefore it is called \it the  birth operator of  the particle with wave function $\Phi$ \rm  and
 denoted $a^*(\Phi )$. So, in the boson case the birth operator is determined with formula
\\$a^*(\Phi ) = \sqrt N\Phi \vee \Psi $\ (13a)\\ and analogous in the fermion case as
\\$a^*(\Phi ) = \sqrt N\Phi \wedge \Psi $\ (13b)
\\ Then we can determine the annihilation operator as a operator conjugated to  $a^*(\Phi )$.
 \it This operator turns the  \rm n- \it particle vector in the \rm (n-1)\it  particle vector.
\\ \rm Such is the briefly  sketch of the second quantization theory. It is important that GNS construction of
every Hilbert space   $\Lambda^ {(m,s)}$   generates the general structure of the Fock space.
\\ Thus there exist the two interdependent approaches to the quantization problem : \it the first
 quantization,\rm which is connected with replacement of Poisson  brackets between canonical
 co-ordinates and canonical momentas  onto the commutation relations with Planck consant $\hbar$. The
 second  approach,
\it the second quantization, \rm is  connected, as we just saw, with the structure of Fock- Hilbert space,
 with the two statistics and  GNS construction.\\
The aim of this article is to show a some original variant of quantization: the quantization of the
\it  particle   velocity. \rm In fact, we will obtain   \it discrete Lorentz groupe. \rm This is one of examples
of  \it quantum \rm  groups,  interest to which is very large now.
\\ Mathematician V.G.Drinfel'd from Kharkow,apparently, is a pioneer in this sphere [9]. One can also
recommend among the  interesting works of this direction the work [10] of G.W.Delius,[11] of P. Schupp,
 [12] of M. Vybornov and many other on this theme.\\
Author wants to   dedicate this work to respectful memory of his first teacher in theoretical physics-
doctor V.M. Pyg, whichdeceased sudden in 1998 y. He was  scientist ,well known with his works in the  sphere of functional integration, conformnal gravitation, fibre bundles both in Soviet Union and far from its frontiers.\\
The invaluable help to author by the investigations, which preceded this work, rendered the employee
 of  Kharkow low temperature Institute- doctor G.N. Geistrin: the well known Soviet and Ukrainian
mathematician. He found the time for me in the difficult conditions of  the 'postsoviet' period in the
 Ukraine; I am very grateful him for this.
\section *{II.The equation of movement for the scalar particle.}
Let us [13] consider the charged scalar spinless particle in the electromagnetic field: the one of most
 simple  theories. We  set here  $\hbar $=c=1.Then the movement equation for this  particle is
 Klein-Gordon-Fock equation
\\$ [(\frac {\partial}{\partial x_m} - i e A_m)^2-m^2 ]\Psi =0 $ (14)
\\ Let us choose the 4-potential $ A_m$ as a \\ $ A_x =A_y =0;A_z=-\frac {E c t }{2};A_0=- \frac {E z} {2}$  \ (15)
 \\ Such choice of potential sets the \it constant electrical field along axis z.\rm We can, without loss of
 generality,work in the two dimensions (z,t). The only Lorentz invariant value:\\$ u= z^2- c^2 t^2$ \ (16)
\\
exists in such space. \\Then we can find  wave function $\Psi (z,t)$ in the form \\$ \Psi (z,t)=G(v) F(u)$ \ (17a)
\\where \\$ v=z+ct$ \ (17b) \\ This decomposition is quite correct.
(The explicit writing down of values c and $\hbar$ helps to see the right dimensions of the  physical values. We will later on, on in the
 case of need, to write down these values explicit).
In particular, the  dimension of  the operator, which acts onto the wave function in (16) is the \it ' square
of momentum'.\\ \rm  Equation  (16)  allows the\it standart Fourier method \rm of solution, i.e. the division of
variables u and v.
\\ It is easy to see that we have for  G(v) the very simple   equation \\$ \frac {v \dot G}{G} = \lambda $ \ (18)
\\ where
 $\lambda$  is the \it division parameter \rm and '$^.$' signifies the derivative by v. The solution of
equation  (18) is \\$ G=$ const $ v^\lambda $ \  (19)
\\ And now let us solve the equation by u, which has the form \\$ u F '' + (1+\lambda ) F ' +
 (\frac {g^2}{16 ( \hbar c)^2}- \frac {k^2}{4}
+\frac {ig \lambda}{4 \hbar c} ) F =0 $ \  (20)
\\ where \\$ k= \frac {mc}{\hbar}$ \ (20a)\\ is \it   Compton wave vector \rm  and \\$ \bf g \rm=e\bf E $\rm  \  (20b)
\\  is the force acted onto  the particle in the electrical field.
\\ One can reduce this equation by means of some transformations to the \it canonical form \rm of the
\it degenerate hypergeometrical equation \rm[14] \\$\theta Z''(\theta) +[(1+\lambda)- \theta] Z'(\theta)-
 (-\frac {A}{\gamma} Z(\theta) =0 $ \ (21)\\ where
\\$ A= (1+\lambda)(\pm i \sqrt C )+ b_2   \\ b_2 =\frac {\lambda ig}{4\hbar c}- \frac {k^2}{4}$ ;
$\gamma = \mp 2i \sqrt C $ \  (21a) \\$ C= \frac {g^2}{16 (\hbar c)^2} $ \ (21b)
\\ The connection of solution Z with solution F is following :\\$ F= exp (H U ) Z $ \ (22)\\ where
 \\$ H= \pm i\sqrt C = \pm \frac {ig}{4\hbar c} $ \ (22a)\\ There exist two linear-independent solutions of
 degenerate hypergeometrical equation (21). The first from them and most important is so-called
 \it Whittaker- Pochhammer series :\\$ Z_1 =R( -\frac {A}{\gamma} ; 1+\lambda ;\theta )=
1+\frac {\theta}{1!} \frac {-( \frac {A}{\gamma})}{1+\lambda}+\frac {\theta^2}{2!} \frac {(-\frac {A}{\gamma})(-\frac {A}{\gamma}+1)}{(1+\lambda )(2+\lambda)}+ ...
\\+ \frac{\theta^n}{n!}\frac {(-\frac {A}{\gamma})(-\frac {A}{\gamma}+1)(-\frac {A}{\gamma}+2)...(-\frac {A}{\gamma}+n-1)}{(1+\lambda )(2+\lambda)...(n+\lambda)}$\rm \ (23)
\\The second   solution has the form {\\$ Z_2 = \theta ^{1-(1-\lambda )} R (-\frac {A}{\gamma}- (1+\lambda)+1; 2-(1+\lambda) ; \theta ) =\theta ^{-\lambda} R (-\frac {A}{\gamma}-\lambda ; 1-\lambda ;\theta ) $ \ (23a)
\\ Let us consider solution (23). The basic demand to the  wave function is its finiteness. That should to
  satisfy this demand we ought to \it cut off  \rm series  (23). It is possible if  \it to set successive equal to
 zero
 $ - \frac {A}{\gamma}$; $  - \frac {A}{\gamma}+1$... \rm Thus we obtain the numerate system of the
wave functions : it  some discrete spectrum. \rm This procedure of cutting off of series (23) is equivalent
 to  \it quantization of field $\Psi$.
\\  \rm Let us consider firstly equation \\$  \frac {A}{\gamma}=0 $ \ (24) \\ Remembering designations
(21) we  obtain the equation for $\lambda$ ( $\lambda$ is the division parameter).
\\$ -\frac {1+\lambda}{2}=(\frac {k^2}{4}- \frac {\lambda i g}{4\hbar c })\frac {1}{\mp 2i\sqrt C}$ \ (25)
\\ We have the two cases; in the fir0st case  (25) has  solution
\\$ \lambda = -\frac {1}{2}- \frac {ip}{2}$ \ (26)\\ where \\$ p=\frac {m^2 c^3}{g\hbar} $ \ (26a)
\\ is the dimensionless parameter.\\ We have in the second case the equation of type
\\$\lambda  0 =t$\\i.e.$ \lambda \to \infty $. This equation has no physical sense.\\ Now we obtain ( with
 account of formulas (22), (26a), (19) )  \it namely the equation, which was proposed in work \rm [12
]:\\$ \Psi (z,t) = exp (i \frac {z^2-c^2t^2}{\hbar c} \frac{g}{4}) (z+ct)^{-\frac {ip}{2}- \frac {1}{2} }$ \ (27)\\ But the situation is more complicated and
interesting  than it see the authors of work  [13]. If we equate
$ -\frac{A}{ \gamma} +1=0$ etc. then we obtain the analogous spectrum
 $ (\lambda_0, \lambda_1,\lambda_2 ...)$ and the corresponding wave functions. Thus solution (27) is
 some \it  zero approach only.\\ \rm Which a  physical sense one can see in above equations? Variable
u (look (16)) has the sense of the \it (two)interval \rm  and the expansion by eigenfunctions $\Psi_n$
means the expansion
 by the   \it  eigenfunctions of square of momentum \rm  (the Fourier image of u  is  square of
 two-momentum).\\ The electrical field $ \bf  E$ \rm  as we know from quantum electrodynamics [6]  is
 the \it source of the ( virtual) electron-positron couples,or the other kinds \\
 of couples .\rm If n is the
 number of such couples then \it is easy to compare  our spectrum  to  such set of
 natural numbers.\\ \rm This allows the simple interpretation in terms of Feynman graphs.The physical
 mass of scalar field $\Psi $ is indeed the  sum of \it its  bare  mass \rm and the \it radiative corrections\rm to the mass,
 which are provoked \it with the  interaction  between the scalar and  the other particles. \rm The
interaction  with the  electron-positron couples is one of such interactions. This is the interaction of
\it electromagnetic type \rm  because field $\Psi $ has the electrical charge. One can construct the\\
effective lines,the vertex operators etc. for such theory. I want to state more in detail the position of the authors
 of work  [13] on function (27) and its connection with  the  birth- annihilation processes in Appendix  . This  position  is very interesting and
important.
\\ But  we ought to consider the one consequence of solution (27)  direct in this section.I quote here
 Yu.P. Stepanovsky and A.S.Bakaj- the authors of work [13]. The question is the
 \it assymptotical behaviour \rm  of  function  $ \Psi (z,t)$ in (27).\\
 The solution (27)  describes the two kinds of particles: with the negative and with the  positive
 charges .Really,potential $A_0$ in (15) becomes infinite and positive by  $ z\to -\infty $ and
 $ t\to \infty $. Hence the  probability for the positive charged  particle to be in the points
 $ z\to -\infty $ is very small. The other situation is at $ z \to -t $ i.e. in the  \it ultra-relativistic region \rm
  near from light cone. Function  $ \Psi (z,t)$  has its maximum in this region.Let us prove this.
\\ Firstly, solution  (19) has some constant. Let us choose as a such constant $\frac {mc}{\hbar}$
 (with the dimension:' length' $^{-1}$ ). Then we can rewrite the second factor in  (27) as a \\$ [(z+ct) \frac {mc}{\hbar}]^{-\frac {1}{2}-\frac {ip}{2} }= e^{(-\frac {1}{2}-\frac {ip}{2} )\rm ln[(z+ct) \frac{mc}{\hbar}]}$
\ (28)
\\(thus \it the index by the exponent is really dimensionless \rm )
\\  At $ z\to -ct \ \rm  ln(z+ct)\to -\infty$.Therefore the  probability for the particle to be in the light cone region
 increases as a \it logarithm.\rm And the particle can to be even in the region $ z \to -\infty$. Issuing from
 the potential distribution (15), we draw a conclusion that
\it  solution \rm (27),\it  especially in the region $ z\to -ct$ \it  describes the particles of both charge signs.
\\ \rm Takes place also the more interesting effect ( although the author of this article don't   knows for
the present ,how to interpret this phenomenon)
\\As it is well known , one can represent the real  logarithm :\\ \rm ln (x)= \rm ln (x+i0y) \\ as a \\ $\rm ln(x) +2k \pi i $
\\
 where $ k \in \bf Z$.\\ \rm Thus \it the infinite degeneracy \rm arises  by the real factors of $ exp (kp \pi) $ type
 in representation  \\$ e^{(- \frac {ip}{2}-\frac {1}{2})\rm ln[\frac {mc}{\hbar} (z+ct)]} =
e^{(- \frac {ip}{2}-\frac {1}{2})\ rm ln[\frac  {mc}{\hbar} (z+ct)]+2k\pi i }$ \ (29)
\\ Thus \it our wave functions acquire in fact the constant extra factors of  $ exp (kp \pi) $ \it type \rm and
we have thus \it the extra spectrum \rm of unknown for the time being nature which is \it  characterized
with the set of quantum numbers \rm k
\section *{III.The Lorentz transformations.}
 Let us consider again solution (27) and let us subject the co-ordinates z and t to  usual Lorentz
transformation \\ $  z'=\frac {z-ivt}{\sqrt{1-\frac {v^2}{c^2}}}$ ;
 $ t' =\frac {z-\frac {ivz}{c^2}}{\sqrt{1-\frac {v^2}{c^2}}}$  \ (30)
\\where v is the velocity of the particle.Then
\\$ z'+ct' = (z+ct) \sqrt{\frac {c-v}{c+v}} $ \ (31) \\and wave function (27) with account of factor
 $\frac {mc}{\hbar}$ and with account of  Lorentz -invariant vector  $\bf E$
\rm ( as it is well known from the theory of electromagnetic \ field [2] , \it  if the direction of  $\bf E $
 \it coincides with axis \rm z \it which is subjected to the Lorentz transformations\rm then this is
correctly)  has form
\\$ \tilde \Psi (z',t') = e^{\frac {-i(z^2-c^2t^2)g}{4 \hbar c}} e^{(\frac {-ip}{2}-\frac {1}{2} )\rm  ln [\frac {mc}{\hbar} \sqrt{\frac  {c-v}{c+v}]} }=
\\ = e^{\frac {-i(z^2-c^2t^2)g}{4 \hbar c}} e^{-\frac {1}{2}\rm  ln [\frac {mc}{\hbar} \sqrt{\frac  {c-v}{c+v}]} }
 e^{-\frac {ip}{2}\rm ln [\frac {mc}{\hbar} \sqrt{\frac  {c-v}{c+v}]} } $\ (32)\\
Since $ e^{-\frac {ip}{2}\ rm ln [\frac {mc}{\hbar} \sqrt{\frac {c-v}{c+v}} (z+ct)]}$
and $e^{-\frac {1}{2}\rm  ln [\frac {mc}{\hbar} \sqrt{\frac  {c-v}{c+v}}(z+ct)]}$ are the terms of solution
(27) that the Lorentz transformation leads to the multiplication of solution (27) onto factor
 $ ^4 \sqrt{\frac{c-v}{c+v}} $ i.e. \it onto the c-number \rm  and hence to the correction of second
exponent.\\ Let us impose the additional demand that \it  second exponent remains invariable by Lorentz
 transformation \rm  (30). This leads to the following equation
\\$ -\frac {p}{2} \rm ln\sqrt{\frac {c-v}{c+v}} =2\pi k ;  k\in (0,\bf Z \rm ^+) $ \ (33)\\ k is the nonnegative
integer number since \it the velocity is positive \rm and p (look (26a)) depends on the sign at
 $\vert \bf g \vert$ \rm  which is '-' in  case of negative charge e.\\ Equation  (33) has the following
solution \\$ V_k = \frac {c[1-e^{-\frac {8 \pi k }{p}}]}{1+ e^{-\frac {8 \pi k }{p}}} $ \ (34) \\Thus \it the
velocities form same discrete spectrum \rm and we now approach object of our work.\\ \bf Exists the
 discrete Lorentz group which leads to the conformnal transformations of the electrical charged field's
 wave function under which therefore the Lagrangian of the theory is invariant.
\\\rm  In fact we obtain \it the quantum Lorentz group .\rm Why draw we suchconclusion? Firstly ,we
 have the numerical set of the permissible velocities.Secondly, the transformation (32) sets
 \it the numerical co-ordinate representation of this Lorentz group  .\rm We can,of course,interpret these
sets of velocities and wave functions  \it as a sets of the Lorentz group  eigenvalues  and the
eigenfunctions  correspondingly. \rm Thus the quantization occurs \it in fact without of Planck
constant  $\hbar$\rm  (compare with article  [10]) .
\\Which are the basic features has this discrete Lorentz group (the fact that it is really a group our reader
 can   prove  easy without our assistance)?\\So, it is \it the discrete subgroup of general Lorentz group ,\rm
which is provoked with the special form of solution (17) of equation (14) with the definite choice of
4-potential (15).This is the \it  Abelian \rm group ,because we have here the rotations in  the one plane. A
s we just this saw,this is[3] \it the discrete subgroup of eigen Lorentz group $ L_+^\uparrow $\rm which is the \it exactly
\rm subgroup of general Lorentz group.
The  Lorentz busts form  the subgroup of   $ L_+^\uparrow$. Our discrete subgroup acts, on the other
 hand, as a \it  Weyl scale reparametrisation \rm multiplying
all wave functions on the constant factor.Thus \it our discrete Lorentz group, which acts as a Weil scale
 reparametrisations group is the subgroup of the busts group. \rm We can interpret our above demand to
 the complex exponential factor in (32)  to be the  Lorentz invariant as a \it gauge fixation which generate
 our discrete  Lorentz group.  \rm Such are the theoretical- group aspects of the obtained discrete
 Lorentz group.
\\ Let us now consider the immovable frame. This frame allows us to observe the movement of  the
 particle with the discrete
set of velocities, i.e. we have \it the set of  eigenvalues  of the velocity operator. \rm Our task now to
construct such operator. Formula (34) \it is the good prompt  in the solution of this problem. \\  \rm Let
us consider  the following function : \\$ f(x) = c \frac {1-e^{-x}}{1+e^{-x}}$  \ (35)\\ and let us  consider
 as a argument x  \it  some operator  $ \hat v $ \rm ( thus we have also \it the function of operator
$  f(\hat v)$) \rm  such that \\$  \frac{8\pi}{p} \hat v =
\frac {8\pi}{p} (c^2 t^2 \frac {\partial^2}{\partial^2 z}-\frac {z^2}{c^2 t^2} ) $ \ (36)\\ It is obvious that
 $\hat v$ is dimensionless and that, because of (34),its  eigenvalues  are the numbers $ k \in [0,\bf Z \rm ^+]$.
Thus we can write down \\$ \frac{8\pi}{p} (c^2 t^2 \frac{\partial^2}{\partial^2 z}-\frac {z^2}{c^2 t^2} ) \tilde \Psi =\frac{8\pi}{p} k \tilde \Psi $
\\or\\$ (c^2 t^2 \frac{\partial^2}{\partial^2 z}-\frac {z^2}{c^2 t^2} )  \tilde \Psi=k \tilde \Psi $  \ (37)
\\for some  eigen-function  $  \tilde \Psi $ of the discrete Lorentz group representation.One can check
that at k=1 $ \tilde \Psi =exp( \frac {-z^2}{2c^2t^2}) $ is \it first  eigen-function of operator $\hat v $.
\\ \rm  Then we find the general solution of equation (37) in form \\$ \tilde \Psi = e^{-\frac {\xi}{2}}
\Theta(z)$ , \ (38) \\where\\ $ \xi= \frac {z}{ct} $ \ (38a) \\ It is easy to see that we so obtain,at the
 minimum formally,\it the equation of  harmonic oscillator
 \\$ (\frac{\partial^2}{\partial^2 \xi}- \xi^2 )\tilde \Psi = k\tilde \Psi $  \rm \ (39) \\ It is the equation of
\it  Schrodinger type \rm and the method of solution of such equations is good described L.D. Landau
and E.M.Lifshitz in their monograph [4].We shall to follow farther this monograph in the statement of our
 method for equations (39).
\\This equation ,in terms of $\xi , \Theta $ has form
\\$ \Theta '' ( \xi)- 2\xi  \Theta '( \xi) +(k-1) \Theta ( \xi) =0 $ \ (40) \\ This is \it   Hermit  equation \rm at
 k-1=2n. But it is easy to show that \it we obtain Hermit polynoms also at \rm  k-1=n \it on the whole set
 $ \bf N$ \it of natural numbers \\ \rm Really, let us denote  k-1= n ; $ 2 \xi =x $ or $ \xi =\frac {x}{2} $ .
 Then we obtain equation
\\$  \Theta '' (\frac {x}{2})- \frac {x}{2}  \Theta '(\frac {x}{2} ) +n \Theta (\frac {x}{2} ) =0 $ \ (41)\\
This is [15]  also the permissible form of Hermit  equation.\\Thus we can  express the  solution  of
equation (41)  as a   \it  family of Hermit polynoms $ H_n$  \rm with account of their normalisation
\\$ \int_{-\infty}^{\infty} e^{-x^2} H_n^2(x) dx= 2^n n!\sqrt \pi $ \ (42) \\and the demand of
 orthonormality for wave functions $  \tilde \Psi $. As a result we obtain
\\$ \tilde  \Psi _n =\frac {1}{\sqrt {2^n n!}} \pi^{-\frac {1}{4}} \frac {1}{\sqrt {ct}} e^{\frac{-z^2}{2c^2t^2}} H_n (\frac {z}{ct}) $  \ (43)\\
Thus we obtained the \it spectrum of operator $ \hat v $ \rm with its values on set
 $ k \in [0,\bf Z \rm ^+]$. But our interest is the \it set  of permissible velocities \rm in form (34), i.e. the
 \it set of  eigenvalues  for the function of  operator $ \hat v $ :
\\$ F(\hat v) =  \frac {c[1-e^{-\frac {8 \pi \hat v }{p}}]}{1+ e^{-\frac {8 \pi \hat v   }{p}}} $\rm  \ (44) \\ It
 is obviously \it  that this function is measurable simultaneously with  numbers \rm  k  i.e.\it the set  of
 functions  $  \tilde \Psi $  \it  is the set  of  eigenfunctions of the velocity operator.
\\ \rm Take place the following equation which is correct by dimension reasons ( we have the
 \it differential operator of  the dimension ' velocity' in left side of this equation \rm ).
\\$ - c^3 t^2 \frac {{\partial}^2 \tilde \Psi }{{\partial}^2 z}=\frac  {c[1-e^{-\frac {8 \pi \hat v }{p}}]}{1+ e^{-\frac {8 \pi \hat v   }{p}}}\tilde \Psi  $ \  (45)
 \\ The discrete nature of our Lorentz group prompts us also the following  step of our investigations.
\\ We  decompose wave  function  $\tilde \Psi  $  in  \it  infinite series
 \\$ \tilde \Psi  = \sum _i c_i \tilde \Psi_i $ \rm \ (46) \\ This is in fact the usual decomposition for the
discrete spectrum in quantum mechanics (look for example [4])). Thus we obtain again the
\it infinite-dimensional Hilbert  space .\rm  By analogy with  \it  three-dimensional Decart  space\rm  we
can say that  $ \tilde \Psi_i $  are the \it orts \rm of the infinite-dimensional   Hilbert  space and
 $ c_i \tilde \Psi_i $  are the
 corresponding  \it projections \rm of wave function $ \tilde \Psi $ onto these 'directions' ( again by
analogy with usual analytical geometry). This is, in general,  the appropriate geometrical interpretation
 for the discrete spectrum in quantum mechanics.\\ It is useful ,for  better understanding of the matter, to
 remember briefly the features of projectors (look, for example monograph [16] ). \\
So, let Hilbert space H disintegrates on direct sum  \\$ H=  G \bigoplus F $  \ (47) \\ of two Hilbert
subspaces G and F  .This means that these spaces are   mutually-orthogonal .Therefore one can represent
 \it  every vector  $ h \in H $  as a \it unique sum \rm
 \\ h=g+f  \ (47a)
\\where  $ g  \in G ; f \in F $ . These vectors g and f are called the \it projections of vector \rm  h
 \it  on subspaces \rm  G  \it and \rm  F    correspondingly. And some operator which does  it is called
 \it the projector. \rm  Let us denote this  operator as P (or $ P_G $ , $ P_F $ when the question is the
projection on the definite subspace). Thus \\$ g=P_G h$  \ (48)
\\ Operator P is  \it  the linear  operator \rm  since Hilbert space is linear. This is also the  \it  limited \rm
operator with the \it unit \rm  norm. Really, \\$ \vert \vert h\vert \vert^2 = \vert \vert g \vert \vert^2 +
 \vert \vert  f\vert \vert ^2 $ \\i.e. \\ $\vert \vert g \vert \vert \leq\vert \vert h\vert \vert $ \ (49)
 \\ or \\ $\vert \vert P\vert \vert \leq 1$ .
\\ But if  $ h\in G $  then g=h , therefore we have the   \it equality \rm in (49). On the other hand, it is well
 known from the theory of operators [17] that  \it  for every limited operator \rm P \it  which acts from the
 one normalised space into the other such space
 \\$ \vert \vert P\vert \vert = \rm sup _{ \vert \vert x \vert \vert \leq 1} \vert \vert Px\vert \vert $ \ \rm (50) \\
 for some vector $ x\in H $. Since we can reach the equality  in (49) then, because of above estimate for
  $\vert  \vert Px\vert \vert $ , operator P has really the unit norm.\\ The  vector g=Ph belongs to space
G by every  $ h\in H $  therefore Pg =g , i.e. $  P^2 H = Ph$. Thus \\$ P^2 =P$ \ (51)\\ Let us consider now
some two  arbitrary vectors $ h_1, h_2 \in H $ and let \\$ h_1 = g_1+f_1 ;   h_2 = g_2+f_2 $  \\Then \\
 $ (g_1, h_2) = (g_1, g_2) =(h_1, g_2) $ \\ i.e. \\ $ (Ph_1, h_2)=(h_1, Ph_2) $\\ for every $  h_1, h_2 \in H
$ . But this means that  \it operator P  \it is the Hermitian   operator \rm \\ P* = P \ (52) \\ And what is
 more :{ \it  this operator is positive} \\$ (Ph,h )\geq 0 $ \ (53) \\ Really \\$  (Ph,h ) =  (P^2h,h )= (Ph,P*h ) = (Ph,Ph )  \geq 0 $
 \\ We can,of course, generalize this theory on the Hilbert space which is the direct sum of the arbitrary
 number of its  subspaces (as we this for example saw in the case of the  multi-particle
 Hilbert space  ). It is obvious also that \\$ \sum _i P_i =1 $ \ (54)\\ for given direct sum of subspaces in
Hilbert space H. The following feature of projectors is also obvious .\\ \it If two arbitrary subspaces
  $ G_1$  \it and  $ G_2$ \it  of Hilbert space\rm   H  \it are mutually- orthogonal then
 \\$ P_{G_1} P_{G_2} =0 $  \rm \ (55) \\ If H is  ortho-normalised  Hilbert space then it is broken up on
the  \it infinite direct sum of its vectors \rm  which corresponds to \it the infinite sum of one-dimensional
 projectors with above features.\\
\rm And let us now consider, how above theory of projectors affects on the  \it theory of self-conjugate
 operators. \rm As we this known (for example from the course of quantum mechanics) such operator
 \it   has the real spectrum .\rm  Our operator of velocity is one of such operators.\\
Let now [3] consider  { \it Hilbert space }  $ \cal H \rm ^2 ( X;\mu)$  \it  of  the complex measurable
 functions with the integrable square and with measure   $ \mu $  \it on space \rm  X.
\\ The every ( real or complex ) measurable function sets ,obviously,  {\ it the operator of multiplication
on} $ \alpha $ \it  on  space \rm  X. Let latter is determined on set  $ D_{\alpha }$. Then we understand
  \it  isomorphism \rm  V  \it  between Hilbert spaces \rm   H \it and  $ \cal H \rm ^2 ( X;\mu)$  \it  with
 feature  $ A= V^{-1} \alpha V$    \it  as a  realisation  of (self-conjugate)  operator \rm  A  \it in Hilbert
 space \rm  H. We also mean that  $ D_{\alpha }= V D_A $ for the set of definition  $D_A $ of operator
A.\\ \it The spectral theorem \rm affirms that  \it every self-conjugate or unitary operator \rm  A  \it in
Hilbert space \rm   H  \it  can be realised with the operator of multiplication in  appropriate Hilbert space \rm
$\cal H \rm ^2 ( X;\mu) $. \\ We  can impart the alternative form to above  theorem in  terms of considered
 us theory of projectors.\\ The feature (54) of  projectors is called  \it  the decomposition of unit \rm in
the theory of projectors .\\ If operator A is realised as an operator of multiplication on the \it  real \rm
function in $\cal H \rm ^2 ( X;\mu) $  then one can introduce  \it some family of the  multiplication
 operators \rm $ e_\lambda $  in $\cal H \rm ^2 ( X;\mu) $  \it of the functions of real parameter  \rm
$ \lambda $  which determined with equalities \\$ e_{\lambda} (x) =1 $ by $ \alpha (x)< \lambda $ \\
and \\$ e_{\lambda} (x) =0 $ by   $ \alpha (x) \ge  \lambda $
\ (56)
\\ It is obviously that  $ e_{\lambda }$ are the  \it projectors \rm in  $\cal H \rm  ^2 ( X;\mu) $ ; therefore
formula \\ $ E_ \lambda = V_{-1} e_{\lambda} V $ \ (57) \\ determines  \it the family of (orthogonal)
 projectors in \rm  H  \it which depends also on \rm  $ \lambda $.\\ It is easy to certain in truth of
following features of projectors:\\
a. $ \ E_\lambda  E_\mu =\ E_\lambda $ by $ \lambda \le \mu $ \  (58a)\\ We obtain formula  (55) as its
 particular case for the  \it orthogonal projectors.\rm \\   b. $\lim_ {\lambda \to -\infty} E_\lambda \Phi =0
$,
 $\lim_ {\lambda \to \infty} E_\lambda \Phi =\Phi $ ,$ \lim_ {\lambda \to \mu-0} E_\lambda \Phi =E_\mu\Phi $ \ (58b)\\ for every $ \Phi \in H $, $\mu \in \bf R  $.
\\ c. The following integral representation takes place (as a  {\it Stieltjes integral})
\\ $ A\Phi = \int_ {- \infty}^\infty \lambda d E_{\lambda} \Phi  $ \  (58c)
 \\ (We want to remind here  our reader that the \it  Stieltjes-( Riemann)  integral of the function \rm f(x)
 \it with the integrable function  \rm g(x)  \it on the (limited) interval\rm  [a,b]  \it is ,by definition,\rm
 \\$ \int {_{a} ^{b} f(x) d g(x)} = \lim _{\max (x_i-x_{i-1})\to 0 } \sum {_{i=1} ^m} f( \xi _i )[g(x_i)-g(x_{i-1}] $
\\where
\\ $ a=x_0< x_1<x_2<...< x_m =b $ \ (a) \\ and $ x_{i-1} \le \xi _i \le x_i $ If g(x) is
\it  some function of limited variation,i.e.  \it  if exists the such positive number \rm M  \it that for division \rm (a) \it of interval
 \rm [a,b]  \it the following equation is true \rm  :\\ $ \sum _{i=1} ^n \vert g(x_i)-g(x_{i-1}) \vert <M $
\\and f(x) \it  is the such continuous function  on \rm  [a,b] \it that above limit exists.\rm One can easy
generalise the definition for  interval  [a,b] on the case \it of the infinite interval, \rm as one makes that in
the usual integral calculus.\\ This theory is quite suitable also for our case of the discrete (\it numerical \rm )
spectrum because of the  limit relations in above definition.\\
The formula (58c) is called the {\it spectral decomposition of self-conjugate operator} A.\\Issuing from
above spectral theory we can now write down the following  definition for the \it function of
 self-conjugate operator \rm A [17] \\ \it  The  function $ \phi (A)$   \it is the operator which is
determined by formula \rm  \\$  \phi (A) f= \int _{-\infty}^ {\infty} \phi (\lambda)dE _\lambda f $ \ (59)
\\ \it  for all vectors $ f \in H $  \it for which formula \rm  (59)  \it is true .\\  \rm And now we can ,in
conclusion, complete our consideration of the discrete Lorentz group theory.\\ We can  \it  compare real
parameter  $ \lambda$ }  \it  to the discrete  spectrum of velocities. \rm In fact ,with accounting of the
 limit nature of formula (58c) in sense of  Stieltjes- Riemann  integral,we have now the following
situation. \\ We showed already that decomposition (46) gets us the representation of some vector
 $\tilde \Psi$  of (ortho-normalised ) Hilbert space H
by its co-ordinates $ \tilde\Psi_i$. Then the substitution of function $\tilde\Psi$  in  formula   (59) and
the comparison of parameter  $ \lambda$  with the set of factors $c_i $ in formula (46)  \it  give  us the
 correct theory for the differential operator of velocity \rm (equation (45))  \it and for its wave function
\rm $\tilde \Psi$.
 \section *{Acknowledgments} I am very grateful G.N.Geistrin and Yu.P. Stepanovsky for their help at
 the investigations preceding this artickle.
I am also grateful D.P. Sorokin for the consultation in the 'technical questions' which were
 necessary for the publication of this article and especially for his moral support of me in my new,at
times difficult, life in Germany.
I am very grateful the scientific society by Jewish community of Germany for the attention rendered my
scientific work.
 \section *{\ Appendix.  The some features of quantization for the scalar field in presence of the homogeneous electric field.}
 We finish our article with the statement of doctors Y.P.Stepanovsky and A.S.Bakaj original point of
 view on the birth -annihilation processes in the scalar field
 electrodynamics. Their theory is connected closely with the theory considered in this article, and it is
very didactic,-to expound it here .\\ So, our careful analysis of solution (27) by choice (15) of 4-potential
 $ A_\mu $  showed that the points with co-ordinates $ z \to -\infty $ are  \it inaccessible \rm at
 $ t \to \infty $ for the particles with the positive charges. And, on the contrary, ultra-relativistic region
$ z \to -ct $ is the region  \it  accessible for the particles of the both sign of charge .\\  \rm The value of
vawe function $\Psi (z,t) $ at $\frac { m^2 c^3}{e E \hbar} >>1$ in above classical (on level of  \it
 classical electrodynamics \rm)  inaccessible region for the  positive charged particles is  \it much less
than its value in the classical accessible points for the  positive charged particles \rm
  $ t \to - \infty ,z \to  \infty $ \\ $ \frac {\Psi (-z,t)}{\Psi (z,-t)} = (-1)^{-i \frac { m^2 c^3}{e E \hbar}}  =(e^{-i\pi}) ^{-i \frac { m^2 c^3}{2e E \hbar}} =e^{-\frac {\pi m^2 c^3}{2e E \hbar}} $ \ (A1)
\\ But latter value becomes  \it  much more than unit \rm  if we adopt  $ -1=exp(i\pi)$ . And this is equal to
 the \it  charge conjugated theory.\\  \rm If we consider wave function (27) as a  function of the complex
variable W= ict+z then one can pass from the negative to the positive z through point W= 0,- \it the
  point of branching out of this  variable.\rm  And we obtain the  \it different \rm  signs in formula
 $ -1=exp(\pm i \pi)$  \it  depend on the direction of the roundabout way of the branching out point. \\
\rm Let us  such modify  function (27)  :
\\ $ \Psi (x,t)= e^{ipx- i{\cal E} t} e^{\pm i\frac {c [(p_z+\frac {eE}{2} t)^2-({\cal E}+\frac {eE}{2}z)^2]}{\hbar eE}} * $
\\$ * [(p_z+ \frac {eE}{2} t) \mp ({\cal E}+\frac {eE}{2} z)]^{\pm \frac {p_x^2-p_y^2+ m^2c^3}{2\hbar eE}-1 } $
 \ A(2)
\\ where $ p_x, p_y,p_z$,$\cal E$ \rm are the values of projection of momentum and energy
 correspondingly;i.e. this function depends on the general 4-vector of momentum in which
homogeneous electric field (15) is joined.Then formula (A1) turns into
\\ $\frac {\Psi (x,y,z,t)}{ \Psi (x,y,-z,t)} =e^{-\pi \frac{p_x^2-p_y^2+ m^2c^3}{2\hbar eE}} $ \ (A3) \\ We
assume now that the electric field is 'engaged' at $ t \to -\infty $ and  'switshed of ' at $ t \to \infty $. This
 field will inevitably generate the \it  charge-conjugate \rm  couples :the scalar bosons of both signs,the
 electron-positron, quark-antiquark and the other couples.\\ We consider now the birth with this field of
the \it  scalare spinless couple of both charges. \rm  And let us consider the such process when
 \it the positive-frequency wave function \rm $ exp(i\bf p \rm _k \bf x-i\cal E \rm _k t) $   \it turn into the
superposition of positive and negative-frequency wave functions \rm  \\ $ e^{i\bf p \rm_k \bf x\rm -i\cal E \rm _k t} \to f  e^{i \bf p\rm _k\bf x \rm -i\cal E \rm'_kt}+ge^{i \bf p\rm _k \bf x\rm +i\cal E\rm _k't} $ \ (A4)\\
 Let  analogous \\ $  e^{i {\bf p} \rm _k {\bf x} \rm  +i{\cal E} \rm _k t} \to f ^*  e^{i {\bf p}\rm _k {\bf x} \rm+i{\cal E }\rm '_kt}+g ^*e^{i{\bf p}\rm _k{\bf x }\rm -i{\cal E} \rm _k't} $ \ (A5)
\\ Then the general solution of equation (14) \it  in  terms of the negative and positive charged paricles
operators \rm  (in Fock space of second quantization) which has the standart form
\\$  \Psi (x,t)= \sum _k \frac {1}{\sqrt{2V{\cal E}_k}}( a_k e^{i \bf p\rm _k\bf x\rm -i{\cal E}'_kt}+b_k e^{i\bf p\rm _k\bf x \rm +i{\cal E}'_kt})$ \ (A6)
\\turns  into\\ $ \Psi '(x,t) = \sum _k \frac {1}{\sqrt{2V{\cal E}_k}}(fa_k+g^ *b^ + _k)e^{i{\bf  p}\rm _k{\bf x}\rm -i{\cal E}'_kt}+(f ^ *b ^+ _k +ga _k)e^{i{\bf p}\rm _k{\bf  x} \rm +i{\cal E}'_kt} $\ (A7)
\\ where $ a_k$ is the annihilation  operator for the particle with momentum $\bf p\rm _k$ and $ b_k^+$ is the birth operator for the particle with momentum $-\bf p \rm _k$.Therefore we can
interpret the factors at the exponent in  (A7) as a correspondingly \it birth and annihilation operators in  electric field \rm  (15):
\\$ a'_k = (f a_k+g^*b_k^+) \sqrt {\frac {{\cal E}'_k}{{\cal E}_k}} $ \ (A8.a)\\ and
\\ $b'_k = (f^*b_k^+ + g a_k) \sqrt {\frac {{\cal E}'_k}{{\cal E}_k}} $ \ (A8.b)
\\ It is obvious that the  \it vacuum average of the operators of number of  partic-\\les \rm
 :$ a'^+ _k+ a'_k$  and $ b'^+ _k b'_k$  \it  is different from zero\rm :
\\$ N_k = <0\vert a'^+_k +a'_k\vert 0> =<0\vert b'^+_k +b'_k\vert 0> =\frac {{\cal E}'_k}{{\cal E}_k}\vert g\vert ^2 $ \ (A9)
\\ where the definition of vacuum (5) , its norm : $ <0\vert 0 >=1 $ and the commutation relations between the birth and annihilation operators  are taken into account.
\\ Formula (A9) shows us that \it  we should to solve classical equation \rm  (14)  \it and to find then
 factor\rm  g  \it from equation \rm  (A4) (this effect of the \it  mixing up of the positive and the negative
 frequencies in  formula \rm (A8) \it is called  Klein para-\\dox \rm ). One can show that this task in  case of
 the homogeneous electric field  'engaged' at $ t \to -\infty $ and  'switshed of ' at $ t \to \infty $ comes to
 the \it  task about the oscillator with the variable frequency.\\ \rm Let us denote the homogeneous
electric field dependence on time as $\bf  E\rm (t)$.This field is generated with the vector-potential
$\bf A \rm (t)$ : $\bf E \rm (t) = -\frac {\partial \bf A }{\partial \rm  t} $ \ (A10)\\ If we represent the
 solution of equation (14) in the form \\ $\Psi (x,t) = \xi(t)e^{i \bf P \bf x} $ \ (A11)\\ then we obtain \it
the  oscillator equation \rm relatively   variable $\xi$:\\ $ \ddot \xi+\omega ^2(t) \xi =0 $ \ (A12)\\ where
\\ $ \omega ^2(t) = (\bf P-\rm e \bf A\rm (t))^2+m^2 $ \ (A13)\\ Let us consider now the case when  field
$ \bf E\rm (t)$ is directed along axis Oz and has the following dependence on time :
\\$ E_z = \frac {E}{\rm ch^2 {\frac {2t}{T}}} $ \ (A14)\\ The vector-potential
\\ $ A_z =-\frac {ET}{2}\rm th \frac {2t}{T} $ \ (A15) \\ The parameter T which we
shall rush to infinity one can interpret as \it  an average time of the action of  field \rm  E , since
 \\$ \int _{-\infty} ^ {-\infty}{ E_z (t)dt} =A_z(-\infty)-A_z (\infty) =Et $ \ (A16)\\ Substituting potential
 (A15) in  (A13) und supposing that $P_z= p_z $ and $ P_y=p_y $ we obtain from equation (A12)\\
 $\ddot \xi +(A \rm th^2 \frac {2t}{T}+B \rm th \frac {2t}{T}+C)\xi =0 $ \ (A17) \\ where
\\ $ A=\frac {(eET)^2)}{4}, B=eETP_z, $\\$ C=p_x^2+p_y^2 +P_z^2 +m^2 $ \ (A17a) \\ This is formally
 equation of oscillator.Let us denote \\$\frac{2}{T} = \varepsilon $ \ (A18)\\i.e. we can rewrite
 equation (A17) as\\ $\ddot \xi +(A th^2  {\varepsilon t}+B th {\varepsilon t}+C)\xi =0 $ \ (A17c)\\
 ($\varepsilon$ has the physical sense \it  of  the energy of oscillator \rm).
\\  It is turn out \it  that one can transform  the such equation of oscillator to  hypergeometrical  equation \rm
\\$ [\eta (\eta-1)\frac {d^2}{d\eta ^2} +((\alpha +\beta +1) \eta-\gamma)\frac {d}{d\eta}+\alpha \beta ]F =0 $ \ (A19) \\where
\\ $\eta=e ^{2\varepsilon t} $ \ (A19a)\\$\alpha =\mu +\lambda +i\frac {\sqrt {A+B+C}}{2\varepsilon} $
\\$ \mu = \pm i\frac {\sqrt {A-B+C}}{2\varepsilon} $,
\\ $ \lambda =\frac {1}{2}\pm \frac {1}{2}\sqrt {-1-\frac {4A}{\varepsilon^2}},
$ \\$\beta = \mu +\lambda -i\frac {\sqrt {A+B+C}}{2\varepsilon}, $ \\$ \gamma =1+2\mu  $ \ (A19b)
\\ (We made here the substitution \\$\xi (t( \eta)) = (-\eta)^{\mu}(1- \eta)^{\lambda}F(\alpha, \beta,\gamma,\eta) $ )
\\ One can represent  solution F as a \it hypergeometrical series\rm
 \\$ F(\alpha, \beta,\gamma,\eta)= 1+\frac {\alpha \bf.\rm  \beta }{1\bf.\rm \gamma} \eta +\frac {\alpha(\alpha+1)\bf.\rm \beta (\beta+1)}{1\bf. \rm \gamma (\gamma+1)} \eta^2 +{\bf...}\rm , $
\ (A20)
\\ If $\alpha$ and $\beta $ are different from 0,-1,-2,...(the serie is broken at these values ) and
$\gamma \neq 0,  -1,-2,...$ then \it  series \rm  (A20)  \it converges absolutely at all \rm
 $ \vert \eta \vert <1 $. According to (A19a) $\eta \to 0$  at $ t\to -\infty$  (we consider parameter
 $\varepsilon$  as a positive parameter).Thus at  $ t\to -\infty$  the solution of equation  (A17c) has the
form \\ $\xi (t(\eta ))=(- \eta) ^ \mu $ \ (A21)\\ (we utilize here the \it majorant \rm estimate for the sum of
 the series).\\We
\it  postulate\rm  now that constants A,B,C in  equation  (A17c) are \it connected with the assymptotical
 values of the oscilator frequencies \rm : $ \omega (\pm \infty) \equiv \omega _\pm  $ and
 $\omega (0) \equiv \omega_0 $ :\\$ A = \frac {\omega_+^2+\omega_-^2}{2}-\omega_0^2, $\\$ B=\frac {\omega_+^2-\omega_-^2}{2}, C=\omega_0^2 $ \ (A22)
\\ Then we can rewrite formula (A21) (with account of denotations (A19 ) ) as
\\$\xi (t(\eta ))=(- \eta) ^{\mu} =e^{\pm i \omega_-t} $ \ (A21a)
\\ Let us ascertain how looks the same solution at $ t \to \infty $ when $ \eta \to \infty $. As we this
known from the theory of hypergeometrical functions[14]
\\$ F (\alpha,\beta,\gamma,\eta) =
\frac {\Gamma(\beta-\alpha)\Gamma(\gamma)}{\Gamma(\beta)\Gamma (\gamma-\alpha)}\eta^{-\alpha} F(\alpha,1-\gamma+\alpha ;1-\beta+\alpha; \frac {1}{\eta})+ $ \\$ + \frac {\Gamma(\alpha-\beta)\Gamma(\gamma)}{\Gamma(\alpha)\Gamma(\gamma-\beta)}\eta^{-\beta} F(\beta,1-\gamma+\beta; 1-\alpha+\beta; \frac{1}{\eta}) $ \ (A23)
\\ According to (A23) at $ t \to \infty $
\\ $\xi (t(\eta )) \approx \frac  {\Gamma(\beta-\alpha)\Gamma(\gamma)}{\Gamma(\beta)\Gamma (\gamma-\alpha)}\eta^{\mu+\lambda -\alpha} +\frac {\Gamma(\alpha-\beta)\Gamma(\gamma)}{\Gamma(\alpha)\Gamma(\gamma-\beta)}
\eta^{\mu+\lambda -\beta} $ \ (A24)\\ or, in  terms of the assymptotical frequencies :
 \\$\xi (t)\approx \frac  {\Gamma(\beta-\alpha)\Gamma(\gamma)}{\Gamma(\beta)\Gamma (\gamma-\alpha)}e^{-i \omega_+t}+ \frac  {\Gamma(\alpha-\beta)\Gamma(\gamma)}{\Gamma(\alpha)\Gamma(\gamma-\beta)}
e^{i\omega_+t} $ \ (A25)\\ The following formula is quite correct
\\$ e^{-i\omega_-t} \to \sqrt {\frac {\omega_-}{\omega_+}} (f e^{-i\omega_+t}+ge^{-i\omega_+t)} $ \ (A26)
\\ We can see,even 'with the naked eye',that this formula is \it fit \rm  for our model of the homogeneous
 electric field 'engaged' at $ t \to -\infty $ and 'switched off ' at $ t \to \infty $. Let this change of the field
is\it  very slow.\rm  We [1] imply under the word ' slow' that
\\ $ T_1 \frac {d \lambda}{dt} \ll \lambda  $ \ (A27)\\ for the field parameter $\lambda $ and the period
$ T_1$.Such slow change of the field is called \it the addiabatical change \rm  of the field. The very
important characteristics of the  \it addiabatical process\ rm  are the values which are \it  invariant \rm  by
 such process. They are called \it the addiabatical invariants.\\  \rm Let us calculate the such
addiabatical invariants for theory (A25) and thus we shall obtain the correct model \it  for the slow
changed  homogeneous electric field.
\\  \rm  We can mark  that the explicit expression for the factors f and g from formula  (A26) we have  from
 formula  (A25).\\ One can represent the arbitrary (real) solution of  oscilator equation at\\ $ t\to -\infty $ as
 \\$ \xi (t)=\sqrt {\frac {2}{\omega_-}} Re (a_-
e^{-i\omega_-t}) $ \ (A28) \\ We can apply  above strategy also to the solution at\\ $ t\to \infty $ and write down \\$\sqrt {\frac {2}{\omega_-}} Re (a_-e^{-i\omega_-t}) \to \sqrt {\frac {2}{\omega_+}} Re (a_-e^{-i\omega_-t}) =$ \\\ $ = \sqrt {\frac {2}{\omega_+}} Re (f a_-e^{-i \omega_+t}+g a_-
e^{-i \omega_+t} =\sqrt {\frac {2}{\omega_+}} Re[f a_-+g*a_-)e^{-i \omega_+t}] $ \ (A29)\\ As we
 known from the course of theoretical mechanics [1] , the integral \\$ I= \frac {1}{2\pi} \oint p dq $ \ (A30)
\\ in the phase space \it is the addiabatical invariant ,\rm  i.e. \\ $ \frac {\bar {dI}}{dt} =0 $ \ (A31)\\  One
 can prove that for the harmonic oscillator with the Hamiltonian
\\$ H = \frac {p^2}{2m} + \frac {m \omega^2 q^2}{2} $ \ (A32)\\ where p - is the momentum of oscillator ;
 m is its mass ; $ \omega $ is its frequency and q is its co-ordinate. It is turns out that
\\ $ I =\frac {E}{\omega} $ \ (A33)\\ where E is the energy of the oscillator. It is easy to see that in our
case \\$
I=\frac {\varepsilon}{\omega} =\vert a\vert^2 $ \ (A34) \\ Really,since $\vert a\vert^2 $ has the
quantum-mechanical sense of \it probability \rm  for  some quantum-mechanical value, then formula
 (A33) means \it that this probability \rm  $\vert a\vert^2 $ \it is in fact constant. \rm Also if one looks
onto  expression (A29) that one can mark that average of $ e^{-i\omega_-t} $  is equal to 1.     Therefore
the addiabatical invariant of the oscillator should to coincide with $\vert a\vert^2 $.\\ Then from (A34)
and (A29) we obtain the formula for the change of the addiabatical invariant I
 \\$ \vert a_-\vert^2 \to  {\vert f a_- +g*a_- \vert}^2 $ \ (A35)\\ Hence
\\ $ \frac {\Delta I}{I} =\frac {I_+-I_-}{I_-} =
\frac { {\vert f a_- +g*a_- \vert}^2}{{\vert a_-\vert}^2} $ \ (A36)\\ since the Wronscian
\\ $ \Psi^* \dot \Psi- \dot \Psi^* \Psi =$ constant \\ for the two complex-conjugate solutions of
 (A28),(A29) then \\$ {\vert f \vert}^2- {\vert g \vert}^2 =1$ \ (A36)\\ Introducing the denotation \\ $ \frac {{\vert g \vert}^2}{{\vert f \vert}^2} \equiv \rho $ \ (A37)\\
and utilising (A36) we obtain \\$ f= e^{i\delta_1}\frac {1}{\sqrt {1-\rho}} $,$g= e^{i\delta_2}\sqrt \frac {\rho}{1-\rho} $ \ (A38)\\ where $ \delta_1$ and $ \delta_2 $ are the phases of f and g correspondingly. Substituting  (A38) into (A36) and representing a as $ a=\vert a \vert e^{i\delta} $ we obtain \\ $ \frac {\Delta I}{I} =\frac
{2 \rm \cos{\phi}  \sqrt \rho+2\rho}{1-\rho} $ \ (A39)
\\ where \\ $ \phi = 2\delta +\delta_1 +\delta _2 $ \ (A39a)\\ Thus \it the exactness of the conservation of
the addiabatical invariant \rm  I \it  depends on the value of \rm  $ \rho $  \it and the  phase
$\phi $ \it of the harmonic oscillation which depends on the  phase \rm  $\delta$
\it at \rm   $ t \to -\infty $. \\ Since the formal equation for the harmonic oscillator
 : $ \ddot x+\omega ^2 tx =0 $  coincides with  \it one-dimensional Schrodinger equation\rm  then it is
 useful to interpret $\rho $ as a \it  factor of the 'over-barrier' reflection \rm  and to utilize in this case {\it the quasi-classical approach}
for the calculation of $\rho $.\\ According to (A25)
\\$ \rho = \vert {\frac {\Gamma (\beta)\Gamma (\gamma-\alpha)}{\Gamma (\alpha)\Gamma (\gamma-\beta)}} \vert ^2 $ \ (A40)
\\ Utilizing the features of $\Gamma $-functions and the fact that  constants
 $\omega _+ ,\omega _ -$ and $ \omega _ 0$ are real, one can find that \\ $ \rho = \frac {\rm ch{ [\pi
(\omega _+-\omega _-)/ \varepsilon ]}+\rm cos (\pi
 \sqrt {1-4A/\varepsilon^2 })}{\rm ch{ [\pi (\omega +-\omega _-)/ \varepsilon}] + \rm cos (\pi \sqrt {1-4A/ \varepsilon^2 })} $ \ (A41)
\\ where, in according with (A22)\\ $ \omega_ {\pm} =\sqrt {A \pm B +C} $
 \ (A41a) \\ The frequency $ \omega (t)$ is not equal to zero if \\$ 4A <{(\omega _+ +\omega _-)}^2 $
 \ (A42) \\  Let us now return to equation (A17). If $ T \to \infty $ and
 $ \vert P_z \vert < \frac { EeT}{2} $ then formula (A41a) yields
\\$ \hbar \omega _{\pm} \approx c [\frac { EeT}{2} \pm P_z + \frac {p_x^2+ p_y^2 +m^2}{eET} ] $
\ (A43)
\\ (we wrote out here the correct dimension' (\it  energy' \rm  ). Then (again in the system $ \hbar =c=1$) we
have the inequality \\$ ( \omega _+ -\omega _-)^2 <4A < ( \omega _++ \omega _-)^2 $ \ (A44)
\\ Utilising (A9) and (A38) we obtain for the average of the operator of  number of particles
\\ $ N_k = \frac {{\cal E}'_k}{{\cal E}_k} {\vert g \vert}^2 \approx \frac {\rho}{1-\rho} $ \ (A45) \\ We
ought now to estimate the value of $ \rho$ .\\ If the condition (A44) is carried out that at $ \epsilon \to 0 $
 we have the following assymptptical estimates for
 $ \rho$  \\  $ \rho = e^{-\frac {2\pi \omega _-}{\epsilon}}  (\omega _- \le \omega _+)  $ ,
\\ $ \rho =e^{-\frac {2\pi\omega _-}{\epsilon}}  (\omega _+ \le \omega _ ) $ \ (A46)\\ if
\\ $ 4A< ( \omega _+-\omega _ )^2 $ , \ (A46a)\\ and
\\ $ \rho = e^{-\frac {\pi[\omega _++\omega _--2 \sqrt {A}] }{\epsilon}} $  \ ( A47) \\ if
\\ $ 4A \ge {( \omega _+-\omega _ )}^2 $ \ (A47a) \\ Then  inequality (A44) leads to estimate (A47a) for $ \rho$  and to formula ( A47).
The substitution of the values of $\omega _+$ and $\omega _-$ in formula ( A47) for $ \rho$  yields the result
\\ $ \rho= e^{-\pi \frac {p_x^2+ p_y^2 +m ^2}{eE}} $ \ (A48)
\\  The values of $ \rho$  are exponential small at $ m^2 \gg eE $,  \it i.e. if  field \rm E \it is addiabatical. \rm
One can neglect the value of
$ \rho$  in  denominator of  (A45) in this case and the value of $ \rho$ { \it defines the average of the borned couples which coincides with the probability of the
 birth of the one couple}\rm because of above trifle of \\ $ \rho$ .One can quote here the general formula
 for the probability of the birth of  the n couples ( the above
probability of the birth of the one couple , -(A48), is the particular case of this general formula):
 \\ $ W_n =(1- \rho)\rho^n $ \ (A49)\\ According to (A49) the exact sense of the value of $ \rho$ is
determined on the following way: the value $ W_0= 1-\rho $ is the probability of the following eve-\\nt
 :\it 'the no couple with the appropriate quantum numbers is borned in the vacuum'.\rm  But as it is well
 known from the course of quantum mechanics ( look, for example,
monograph [4], paragraph 41) \it the  quantum- mechanical state with the definite quantum numbers is
(in fact) the invariant by the addiabatical evolution of the quantum system . \rm This statement, which
was proved with the help
of the mathematical apparatus of the { \it quantum perturbative theory} is called{ \it   P.Ehrenefest
hypothesis}.  Thus, in   terms of  P.Ehrenefest hypothesis, the value of $ \rho$  determines
\it non-fulfilment of  P.Ehrenefest hypothesis.
\\ \rm  Let us find { \it the full  number of  couples which are  borned with electric field }E \it at  \rm
 $ m^2 \gg eE $. Let us multiply for this the value of $ \rho$  onto the element of the phase volume
 $ V d p_x d p_y d P_ z / (2{\pi \hbar})^3 $ and let us
integrate over the all permissible values of momentum ($ \rho$  has the sense \it of the density of
probability \rm ).
It is necessary to integrate over $ d P_ z$in the limits from - EeT/2 to EeT/2 . It is connected with the truth
of  formula  (A48)   for  $\rho $  at  $ \vert P_z \vert< \frac{eET}{2}$
. The analysis of formula (A47) at $ \vert P_z\vert >\frac {eET}{2} $and $ T \to \infty$ shows us that
$\rho =0 $ in this case. After the integration ( as a Gauss integral)  we obtain
\\ $ N= \frac {V}{{(2\pi \hbar)}^3}\int e^{-\pi  \frac
{p_x^2+p_y^2+m^2}{eE}}dp_x dp_y dP_z = $ \\ $ = VT  \frac {{(eE)}^2}{{(2\pi \hbar})^3 }e^{\frac {-\pi m^2}{eE}} $ \ (A50)
\\\\ Thus \it the average number of the couples
borned with the addiabatical changed homogeneous electric field \rm E  \it in the unit of the volume
during
the  unit of  the time \rm is \\ $ n=\frac {{(eE)}^2}{{(2\pi \hbar})^3 }e^{\frac {-\pi m^2}{eE}} $ \ (A51)
\section*
\refname
 { \rm [1] L.D. Landau and E.M.Lifschitz ,\it  Mexanika\rm ,Moskow,
'Fizmatgiz' (1953).}\\{[2] L.D. Landau and E.M.Lifschitz ,
{\it Teorija Polja}, Moskow,
 'Fizmatgiz' (1960).}\\{[3] N.N. Bogolubov,A.A.Logunov,A.I.Oksak,I.T.Todorov, \it Obschtie
Printsipy Kvantovoj Teorii  Polja,\rm Moskow,'Nayka' (1987).}\\{ [4]L.D. Landau and E.M.Lifschitz ,
\it Kvantovaja Mexanika. Nereljativistskaja  Teorija\rm (the 4$^{-th}$edition of L.P. Pitaevsky ),
\rm Moskow,'Nayka'
(1989).}\\{[5] V.G. Levich, J.A. Vdovin, V.A. Mjamlin, \it Kyrs  Teoreticheskoj Fiziki, \bf v.\rm
2,Moskow,'Nayka'
(1971).}\\{ [6] A.I. Akhiezer and V.B. Berestetsky, \it Kvantovaja Elektrodinamika,\rm Moskow,'Nayka'
 (1969).}
\\{[7] Lewis H. Ryder ,\it Quantum Field Theory,\rm  Cambridge University Press (1984).}
\\ {[8]  R. Penrose and W. Rindler ,\it Spinors and Space-Time,\bf v.\rm 1, \it Two-Spinor Calculus and
Relativistic
Fields, \rm Cambridge University Press (1984).}
\\{[9]  V.G. Drinfel'd,\it Hopf algebras and the quantum Yang-Baxter
equation,\ rm Sov.
Math. Dokl. \bf 32\rm (1985) p.254.}\\{[10]  Gustav W. Deliusy,
\it Introduction to Quantum Lee Algebra, \rm q-alg$\setminus$9605026.} \\
{[11] Peter Schupp, \it Quantum Groups, Non-Commutative Diff
erential Geometry
and Applications\rm , hep- th$\setminus$9312075.}
\\ {[12]Maxim Vybornov,\it On Quantum Jakobi Identity,\rm q-alg$\setminus$9607007.}
\\{[13] A.S. Bakaj and Ju.P.Stepanovsky,\it  Addiabaticheskie invarianty,\rm Kiev, 'Naukova Dumka'(1981).}
\\
{[14] H. Bateman and A. Erdelyi,\it Higher Transcendental Functions ,\bf v.\rm 1, Mc Graw-Hill Book
Company,Inc(1953).}
\\{[15]   E. Jahnke and F. Emde, \it Tables of functions with formulae and
curves, \rm  Dover Publications, Inc., New York, (1945).}
\\{ [16] N.I. Akhiezer and I.M. Glasman, \it Teorija Linejnyx Operatorov V Gilbertovom Prostranstve,
\bf v.
\rm 1, the publising house at Kharkov State university (1977)} \\
{[17] A.N. Kolmagorov and S.V. Fomin, \it Elementy  Teorii Fynktsij I Fynktsional'nogo Analiza,
\rm Moskow,'Nayka' (1976).}
\end{document}